\documentstyle[12pt]{article}
 \textwidth
 16.4cm
 \oddsidemargin
 2.5cm
 \advance\oddsidemargin
 by
 -1in
 \evensidemargin
 0.0cm
 \advance\evensidemargin
 by
 -1in
 \marginparwidth
 1.9cm
 \marginparsep
 0.4cm
 \marginparpush
 0.4cm
 \topmargin
 -1.5cm
 \advance\topmargin
 by
 -0.0in
 \textheight
 23.0cm
 \makeindex

 \newcommand{\doublespace}
 {
 \renewcommand{\baselinestretch}
 {1.6}
 \large\normalsize}
 
 \newcommand\ra{\rangle}
 \newcommand\noi{\noindent}
 \newcommand\beq{\begin{equation}}
 \newcommand\eeq{\end{equation}}
 \newcommand\beqn{\begin{eqnarray}}
 \newcommand\eeqn{\end{eqnarray}}
\begin{document}
\vspace*{1cm}
\hspace*{9cm}{\Large MPI
H-V29-1998}
%\\
%\hspace*{9.5cm}{\Large
%nucl-th/98xxxxx}

\vspace*{3cm}

\centerline{\Large
{\bf
Neutron-Antineutron
Oscillations in
Nuclei Revisited}\footnote{To be published in proc. of
the
{\it Workshop on Future Directions
in
Quark Nuclear Physics}, Adelaide, March 9 - 20,
1998}}

\vspace{.5cm}
\begin{center}
 {\large
J\"org~H\"ufner$^{1,2}$
and
Boris~Z.~Kopeliovich$^{2,3}$}\\
\medskip
{\sl $^1$Institut f\"ur Theoretische
Physik
der
Universit\"at,
Philosophenweg
19,
\\
69120
Heidelberg,
Germany}\\

{\sl
$^2$
Max-Planck
Institut
f\"ur
Kernphysik,
Postfach
103980,
%\newline
69029
Heidelberg,
Germany}\\

%\vspace{0.3cm}

{\sl
$^3$
Joint
Institute
for
Nuclear
Research,
Dubna,
141980
Moscow
Region,
Russia}

\end{center}

\vspace{.5cm}
%\doublespace
\begin{abstract}

An upper limit on $n-\bar n$ oscillations can be obtained
from
the
stability of matter. This relation has been worked
out
theoretically
and together with data yields $\tau_{n\bar n} > (1 -
2)\times
10^8\,sec$.
A recent publication claims a different relation and finds
a
nuclear suppression of the $n\to\bar n$
oscillations
which is two orders of magnitude weaker than the
previous
evaluations. Using the same approach 
we find, nevertheless, that the earlier estimates are
correct
and conclude that future experiments with free neutrons
from
reactors
are capable to put a stronger limit on the $n - \bar
n$
oscillation
time than experiments with large amount of neutrons bound
in
nuclei.

\end{abstract}

\bigskip

\newpage

\doublespace

\noi
{\large\bf
1. Introduction}
\medskip

Although the proton decay expected in a simple version
of
grand
unified
theories has not been observed with a large lower limit
on
the
lifetime,
other
variants of the theory predict reactions with nonconservation of
baryon minus lepton number by two units,
like
neutron-antineutron oscillations
\cite{kuzmin}
or neutrinoless double-$\beta$ decay (see
e.g. \cite{d-beta}).
The former
phenomenon
has been
searched
for
with free neutrons from reactors in the recent
experiment
ILL-Grenoble
\cite{ill} and the lower limit for the
oscillation
time
\beq
\tau_{n\bar n} > 0.86\times
10^8\,sec
\label{1}
\eeq
\noi
is established. A new experiment planned at ORNL
\cite{yura}
is
intended to increase this limit by two orders
of
magnitude.

Experiments which search for the proton decay are sensitive
to
$n
\to \bar n$ transitions as well, since an antineutron produced inside
a
nucleus
annihilates and blows the nucleus up. Such events are
not
observed
above
the
background in the
experiments
\cite{kamiocande,frejus}.
The experiments \cite{kamiocande,frejus} established
quite
high
lower limits on the lifetime of the
nuclei,
respectively,
\beq
T(O^{16}) > 4.3\times
10^{31}\,yr\\
\label{2}
\eeq
\beq
T(Fe^{56})
> 6.5\times
10^{31}\,yr
\label{3}
\eeq
\noi
These data can be used to deduce a limit on
$\tau_{n\bar
n}$
provided
the nuclear effects
are
known.
Two groups \cite{dgr} and \cite{wanda}
have calculated
the
nuclear effects and based on these data
have provided close
estimates
of
\beq
\tau_{n\bar n} > (1 - 2)\times
10^8\,sec\
.
\label{4}
\eeq
\noi
Recently, these calculations have been challenged \cite{lenya}
and
via
the nonrelativistic diagram technique a value of $\tau_{n\bar
n}$
is
deduced from Eqs~(\ref{2})-(\ref{3}), which is about two
orders
of
magnitude higher than Eq.~(\ref{4}).  If the latter
calculation
were
true, this lower limit
would
be
comparable to what is planned
for
future
experiments with free
neutrons
\cite{yura}.

In this note we discuss this controversy and
conclude
that
the theory which leads to (\ref{4}) is correct. We
suggest
an
intuitive physical picture of a possible $n\to\bar n$ transition
in
a
medium and point out why the approximation used
in
\cite{lenya}
is not valid. In our calculations
we
emphasize
that at low energy the
annihilation
radius
is quite large which may
lead
to
substantial corrections
to
(\ref{4}).\\
\\

\noi
{\large\bf
2. Interference effects in $\boldmath n-\bar
n$
oscillations}
\medskip

We assume that the physical
neutron
(antineutron),
{\sl i.e.} the
vacuum
eigenstate,
contains an admixture of a $\bar n$
($n$)
component.
Suppose that at time $t=0$ we have a pure
$|n\rangle$
state
(neutrons from a reactor, see below), which is a mixture of
the
physical
$|n\rangle_{phys}$ and $|\bar n\rangle_{phys}$
having
different
masses. As a result of a phase shift between these
states
which
increases with time a $|\bar n\rangle$ component appears with
a
probability
proportional to the square of time (compare with 
the well known
effect
of
$K^0 - \bar K^0$
oscillations),

\beq
W_{\bar n}(t)
\approx
(\epsilon\,t)^2\
\label{6}
\eeq

\noi(provided
that
$t\ll1/\epsilon$). Here $\epsilon=1/\tau_{n\bar n}$ is related to the
off diagonal matrix element of the vacuum Hamiltonian, 
and $\tau_{n\bar n}$ is the $n\to\bar n$ oscillation time.

The quadratic time dependence is a direct manifestation
of
quantum
coherence, a linear dependence would be the result
in
classical
physics. This
fact
is extremely important for an understanding of
the
suppression
mechanism
$n \to \bar n$ transitions in nuclei. Let us consider a set up
where
a
neutron propagates a distance $L$ to
a
counter
which can measure the antineutron component. In another set
up
we
have $N$ such counters each separated by a
distance
$L/N$.
Each counter filters away the
antineutron
component.
In the first case we have a probability $W^{I}_{\bar
n}
\approx
(\epsilon\,L/v)^2$, in the second set up $W^{II}_{\bar
n}
\approx
N\,(\epsilon\, L/Nv)^2 \approx W^{I}_{\bar n}/N$, to detect
the
$\bar
n$.Here $v$ is the velocity of the neutron. The second case
is
closely
related to the absorption in a nucleus, where each
bound
nucleon
serves as a detector for the $\bar
n$
component.\\

{\bf 2.1 The mean lifetime of an antineutron in
a
medium}
\medskip

The mean free time of propagation of the antineutron
through
nuclear
matter
(without
annihilation)
is
$t_{free}=1/(\sigma_{ann}v\rho)$,
where $\sigma_{ann}$ is the $\bar n - N$ annihilation
cross
section
and $\rho$
is
the
density of nucleons. The smallness
of
$t_{free}$
is one important
source
of
suppression of $n \to \bar n$
transitions
in
a medium. A neutron which propagates through
a
nuclear
medium produces an antineutron which annihilates with
a
rate
\beq
\frac{dW_{\bar
n}}{dt}
=
\sigma_{ann}v\rho\
\epsilon^2
t_{free}^2\
,
\label{7}
\eeq

In the limit of small velocity one can replace
$\sigma_{ann}$
by
the imaginary part of the $\bar n - N$ scattering
length
\cite{lenya},

\beq
v\sigma_{ann} = \frac{4\pi}{m}\ {\rm
Im}\,a_{\bar
nN}
\label{8}
\eeq

\noi
The annihilation rate (\ref{7}) turns
out
to
be {\it inversely proportional} the
scattering
length
\beq
\frac{dW_{\bar
n}}{dt}
=
\frac{m\epsilon^2}{4\pi\rho\, {\rm Im}\,a_{\bar
nN}}\
,
\label{9}
\eeq
\noi
This result is quite different from that in \cite{lenya},
where
the
annihilation rate is found to be {\it proportional} to ${\rm
Im}\,a_{\bar
nN}$.
Below we show that this might be true only in
the
limit
of very small density $\rho$ which is not the case for a
nuclear
medium.\\

{\bf 2.2 The coherence and
formation
times}
\medskip

There is another source of the $n \to \bar
n$
suppression.
In the presence of an external
field
which
acts differently on $n$ and $\bar n$, the transition $n \to
\bar
n$
is forbidden by energy
conservation. It
may
happen only virtually, with a lifetime
of
the
$\bar n$ fluctuation in
the
neutron,

\beq
t_c = \frac{1}{\Delta
E}\
,
\label{10}
\eeq

\noi
where $\Delta E$ is the energy splitting between $n$ and
$\bar
n$.
In order to be detected the $\bar n$ fluctuation
has
to
interact (annihilate) during
its
lifetime.

Note that at the far periphery of a nucleus the neutron is
quasi-free,
{\it i.e.} $\Delta E \to 0$, and one may conclude from
(\ref{10})
and (\ref{9}) that the $\bar n$ fluctuation lifetime
has
no restriction. This is not true because the neutron cannot
stay
at the periphery longer than $1/\omega$, where $\omega$ is
the
oscillator frequency (the shell
splitting).
Thus, one must
replace
effectively $\Delta E \to \Delta E + \omega$. This is equivalent
to
inclusion of the formation time effects
\cite{hk}.

We conclude that the effect of coherence and formation 
time becomes the main
source
of $n \to \bar n$ suppression in the case of low
density,

\beq
\frac{dW_{\bar
n}}{dt}
=
\frac{4\pi\ \rho\ {\rm
Im}\,a_{\bar
nN}}
{m\
(\Delta
E+\omega)^2}
\label{11}
\eeq

The (unrealistic) limit of low density is recovered
by
calculations
in \cite{lenya}. There the energy
splitting
between
$n$ and $\bar n$ in nuclear medium
is
taken
to be $\Delta E = 8\
MeV$,
the
mean binding energy of a bound nucleon. It is assumed that
the
antineutron
appears momentarily and is
unable
to
feel the mean field of the whole nucleus, therefore,
it
has
no binding. This cannot be true because the $\bar
n$
fluctuation
lifetime (\ref{10}) is very long in this case,
$t_c=25\
fm$. \\

{\large\bf 3. The evolution equation for a
bound
neutron}
\medskip

To incorporate all the above effects one should
consider
the
evolution of a $n - \bar n$ system in
a
medium.
This is a typical two-channel problem (see
e.g. in
\cite{hk,dgr}).

\beq
i\frac{d}{dt}\left|\Phi(t,\vec
r)\right\ra
=
\hat H(\vec r) \left|\Phi(t,\vec
r)\right\ra\
,
\label{12}
\eeq

\noi
where the wave function $|\Phi(t,\vec r)\rangle$
contains
the
two components $n$ and $\bar n$ and depends on time $t$
and
the
space $\vec r$. The evolution operator $\hat
H(\vec
r)$
has
a
form,

\beq
\hat H(\vec
r)
=
\left(\begin{array}
{cc}E_n(\vec r)&\epsilon\\\epsilon&E_{\bar
n}(\vec
r)-
{2\pi i\over m}\widetilde\rho(\vec r){\rm
Im}\,a_{\bar
nN}
\end{array}\right)
\label{13}
\eeq

\noi
Here $\widetilde\rho(\vec r)$ is the nuclear
density
modified
due to annihilation radius $R_{ann}$ which may
be
large,

\beq
\widetilde\rho(\vec
r)
=
\left(\frac{3}{2\pi
R^2_{ann}}\right)^{3\over
2}
\int d^3s\ \rho(\vec
r+\vec
s)\
\exp\left(-\frac{3s^2}{2R^2_{ann}}\right)
\label{13a}
\eeq

\noi
The modified density $\widetilde\rho(\vec r)$
spreads
to
larger distances than
$\rho(\vec
r)$.

The solution of equation (\ref{12}) for the weight of the
antineutron
component
at long times ($t \gg t_c,
t_{free}$)
reads,

\beq
W_{\bar
n}(t,r)
=
\frac{\epsilon^2}
{[\Delta E(\vec
r)+\omega]^2
+
((2\pi/m)\widetilde\rho(\vec r){\rm
Im}\,a_{\bar
nN})^2}
\label{14}
\eeq

\noi
We corrected here for the oscillator frequency considered
in
the previous
section.

One should average expression (\ref{14}) over
different
neutrons
which have different binding energies dependent on a
shell
\cite{dgr}.
However, this difference is much smaller that the
binding
energy
of an antineutron used in \cite{dgr,wanda} which was assumed
to
be
independent of the shell. Therefore, one can safely use the mean
value
of
$\Delta E$ in (\ref{14}) and then sum over
the
neutrons.

The annihilation rate of antineutrons produced in a
nucleus
is
given by integral of (\ref{14}) over the
nuclear
volume
weighted by neutron density and annihilation
cross
section,

\beq
\frac{\Gamma_A}{N}
=
\frac{\epsilon^2}{T_R}=
\frac{m\,\epsilon^2}{A\,\pi\,{\rm
Im}\,a_{\bar
nN}}
\int d^3r\ \frac{\rho(\vec
r)}{\widetilde\rho(\vec
r)}
\
\frac{1}{1+K^2(\vec
r)}
\label{15}
\eeq

\noi
Here $A$ and $N$ are the atomic and
neutron
numbers
of the nucleus,
respectively,
and

\beq
K(\vec
r)
=
\frac{\Delta
E(\vec
r)+\omega}
{(2\pi/m)\,\widetilde\rho(\vec r)\,|{\rm
Im}\,a_{\bar
nN}|}
\label{16}
\eeq

The assumption made in \cite{lenya} corresponds
to $K(r)\gg 1$. This could be not a very rough 
approximation, but $\Delta E=8\,MeV$ used in~\cite{lenya}
is an order of magnitude smaller than in~\cite{dgr,wanda}
and here (see discussion at the end of previous section). 
This is the main reason why $\Gamma_A$ is
overestimated by two orders of magnitude in~\cite{lenya}.\\

{\large\bf
4. Numerical
evaluation}
\medskip

Following to \cite{dgr} we assume that the $\bar
n$
binding
energy depends on $r$ in the same way as
$\rho(\vec
r)$.
Therefore, (\ref{16}) is approximately a constant
$K(r)\approx
K(0)$
at $r < R_A$, where $R_A \approx r_0\,A^{1/3}$ is the
nuclear
radius.
The ratio $\rho(\vec r)/\widetilde\rho(\vec r)$ in
(\ref{15})
is
unity at $r < R_A$, but then steeply
falls
down. Using
these observation we can perform a very simple (but
rough)
evaluation
of
(\ref{15}),

\beq
\frac{\Gamma_A}{N}
\approx
\frac{4\,m\,\epsilon^2\,r_0^3}{3\,{\rm
Im}\,a_{\bar
nN}\,
(1+K^2(0))}
\label{17}
\eeq

\noi
A spectacular observation which follows from this result
is
$A$-independence
of the nuclear annihilation width. 
Of course,  better calculations (see below)
may lead to a weak A-dependence.
The claim in
\cite{dgr} that annihilation rate gets main
contribution
from
the neutrons on the surface might be misleading. In
fact
their
result is also
nearly
$A$-independent,
what supports our observation that all the nuclear
volume
contributes
(see also
in
\cite{lenya}).

The $\bar pp$ scattering length
was
determined
in \cite{sl} using the LEAR data on $\bar
pp$
atoms,
$|{\rm Im}\,a_{\bar pp}| \approx
0.7-1.2\
fm$. Following
\cite{lenya} we use the same value for $\bar
nn$
and
$\bar np$. With this value
we
have,

\beq
\frac{2\pi}{m}\,\widetilde\rho(0)\,
|{\rm
Im}\,a_{\bar
nN}|
\approx 30-50\
MeV\
,
\label{17a}
\eeq

\noi
where we use the central density
$\widetilde\rho(0)=0.16\
fm^{-3}$.

Expression (\ref{17a}) corresponds to the imaginary part of
the
antineutron nuclear potential in \cite{dgr}.  The real part of
the
antineutron potential in the center of the nucleus can be only
guessed
since the data on antineutron atoms are sensitive only to the
surface
of the nucleus.  This is the reason of diversity of
antineutron
potentials used in \cite{dgr,wanda}, which we consider as
very 
unreliable.  An advantage of \cite{lenya} and our approach
is
that we use experimental information on $n\bar n$ scattering
length,
rather than guess a value of imaginary part of the nuclear
potential.
For a simple evaluation we choose the real part to be
of
the same order as the imaginary one, {\it i.e.} that $K(0)\approx
1$ ($\Delta E(0) \approx 40\,MeV$).
This leads to the estimate for the nuclear factor $T_R$ in
(\ref{15}), $T_R \approx 0.23\,fm^{-1}$,
which is quite close to the results
of
exact integration in (\ref{15}):
$T_R=0.20\,fm^{-1}$ for $^{16}O$
and
$T_R=0.18\,fm^{-1}$ for
$^{56}Fe$.
For these
calculations
we used the oscillator parameter $\omega$
from
\cite{molinari} and $40\,MeV$ for
(\ref{17a}).
Following \cite{dgr,wanda} we assumed that $\Delta
E(r)
\propto \rho(r)$. We use the realistic Woods-Saxon
parameterization for the nuclear density.
Dependence of the result on the annihilation
radius
turns out to be quite weak. We use
$R_{ann}=1\,fm$.

We take the upper bound for $\Delta E(0)$ to be
$200\,MeV$
which exceeds most of model values used in
\cite{dgr,wanda}.
Corresponding values of $T_R$ are $1.2\,fm^{-1}$ for
$^{16}O$
and $1.3\,fm^{-1}$ for
$^{56}Fe$.

This estimation for nuclear effects combined
with
the experimental limits for the lifetime of the
nuclei
leads to the following restrictions for the oscillation
time
$\tau_{n\bar
n}=1/\epsilon$,

\beqn
\tau_{n\bar n}(^{16}O)
&>&
(0.6 - 1.5)\times
10^8\,sec\\
\tau_{n\bar n}(^{56}Fe)
&>&
(0.75 - 1.9)\times
10^8\,sec
\label{19}
\eeqn

\noi
These results are very close to the low bound (\ref{1}) found 
in experiment \cite{ill} with free neutrons and are compatible
with (\ref{2}) - (\ref{3}) as well.

The uncertainty of the lower 
limit (\ref{19}) is still
conventional
since it depends on the guessed real part of the
antineutron
potential inside the
nucleus.

Concluding, we do not see how the lower limit for
$\tau_{n\bar
n}$
can be pushed up significantly by using the stability
of
matter
and one may have to turn again to experiments with
free
neutrons
with long flight path and
small
velocity.

{\bf Acknowledgements:} We are grateful to Yuri Kamyshkov 
who encouraged us to write this paper for useful comments,
and to Avraham Gal and Leonid Kondratyuk for
clarifying discussion. 
B.Z.K. is thankful to Tony Thomas for hospitality at CSSM
where a part of this work was done, and for a 
financial support.

\end{document}